\documentclass[a4paper,11pt]{statsoc}
\usepackage{amssymb}
\usepackage{graphicx}
\usepackage{natbib}

\title[Ratings and rankings]{Ratings and rankings: Voodoo or Science?}
\author{Paolo Paruolo\thanks{{\it Address for correspondence:} Paolo Paruolo,
Department of Economics, University of Insubria,
Via Monte Generoso 71, 21100 Varese, Italy. \\ E-mail: paolo.paruolo@uninsubria.it \\
{\it  Date:} \today}}
\address{Department of Economics, University of Insubria, Varese, Italy.}
\email{paolo.paruolo@uninsubria.it}
\author{Michaela Saisana}
\address{European Commission, Joint Research Centre, Ispra (VA), Italy.}
\author[Paruolo, Saisana \& Saltelli]{Andrea Saltelli}
\address{European Commission, Joint Research Centre, Ispra (VA), Italy.}

\begin{document}

\begin{abstract}
Composite indicators aggregate a set of variables using weights which
are understood to reflect the variables' importance in the index.
In this paper we propose to measure the importance of a given variable
within existing composite indicators via Karl Pearson's `correlation ratio';
we call this measure `main effect'.
Because socio-economic variables are heteroskedastic and correlated,
relative nominal weights are hardly ever found to match relative main
effects; we propose to summarize their discrepancy with a divergence
measure. We discuss to what extent the mapping from nominal
weights to main effects can be inverted. This analysis is applied to six
composite indicators, including the Human Development Index and
two popular league tables of university performance.
It is found that in many cases the declared importance of single indicators
and their main effect are very different, and that the data correlation structure
often prevents developers from obtaining the stated importance, even
when modifying the nominal weights in the set of nonnegative numbers
with unit sum.
\end{abstract}

\email{paolo.paruolo@uninsubria.it}

\section{Introduction}
\label{sec_Introd}
In social sciences, composite indicators
aggregate individual variables with the aim to capture
relevant, possibly latent, dimensions of reality such as a country's
competitiveness (\cite{wef2010}), the quality of its governance (\citet{ROL2010}),
the freedom of its press (\cite{REF2011,FH2011}) or the efficiency of its universities
or school system (\citet{LeckieGoldstein2009}).
These measures have been termed `pragmatic' (see \citet{Hand2009}, pp.
12-13), in that they answer a practical need to rate individual units (such as
countries, universities, hospitals or teachers) for some assigned
purpose.

Composite indicators (which are also referred to here as indices) have been increasingly adopted by
many institutions, both for specific purposes\ (such
as to determine eligibility for borrowing from international loan programs)
and for providing a measurement basis for shaping broad policy debates,
in particular in the public sector (\citet{Bird2005}).
As a result, public interest in composite indicators has enjoyed a fivefold
increase over the period $2005-2010$: a search
of  `composite indicators' on Google Scholar gave $992$ matches on
October $2005$ and $5,340$ at the time of the first version of this paper (December $2010$).

Composite indicators are fraught with normative assumptions
in variable selection and weighting. Here
`normative' is understood to be `related to and dependent upon a system
of norms and values'. For example, the proponents of the
Human Development Index (HDI) advocate replacing
gross domestic product (GDP) per capita
as a measure of the progress of societies with a
combination of (i) GDP per capita (ii) education and (iii)
life expectancy, see \citet{Ravallion2010}.
Both the selection of these three
specific dimensions and the choice of building the index
by
giving these dimensions equal importance
are normative,
see \citet{Stiglitz2009}, p. 65.
Composite indicators are thus often the subject
of controversy, see \citet{Saltelli-2007}, \cite%
{Hendrik-etal-2008}.

The statistical analysis of composite indicators is
essential to prevent media and stakeholders taking  them at
face value (see the recommendations in \citet{OECD2008}), possibly leading to
questionable policy choices. For example, a policy maker might think of merging higher education
institutions just because the most popular league table of universities puts a prize on
larger universities, see \citet{Saisana-al11}.

Most existing composite indicators are linear, i.e. weighted arithmetic averages \cite[]{OECD2008}. Linear aggregation rules have been criticized
because weaknesses in some dimensions are compensated by
strengths in other dimensions; this characteristic is called `compensatory'.
Non-compensatory and non-linear aggregate ranking rules
have been advocated 
by the literature on multicriteria decision
making, see for example
\citet{Billaut-etal-2010}, \citet{Munda2008}, \citet{MundaNardo09},
\citet{Balinski2010}.
In this paper we concentrate on linear aggregation, because
of its widespread use.

In this paper we address the issue of measuring variable importance
in existing composite indicators.
As illustrated by a motivating example at the end of this section,
\textit{nominal weights are not a measure of variable importance},
although weights are assigned so as to reflect some stated
target importance, and they are communicated as such.
In linear aggregation, the ratio of two nominal
weights gives the rate of substitutability between the two individual variables,
see \cite[Chapter 4]{Bouyssou2006}, or \citet{Decancq-Lugo2010},
and hence can be used to reveal the target relative importance of individual indicators.
This target importance can then be compared with ex-post measures of variables' importance,
such as the one presented in this paper.

We propose to measure the importance of a given variable via
Karl Pearson's `correlation ratio', which is widely applied in global sensitivity analysis as a first-order sensitivity measure; we call
this measure `main effect'. Main effects represent the expected relative
variance reduction obtained in the output (the index)
if a given input variable could be fixed
(\cite{SaltTara02}, see Section 3.1). They are
based on the statistical modelling of the relation between the variable and
the index.

This statistical modelling can be parametric or non-parametric;
we compare a linear and a non-parametric alternative based
on local-linear kernel smoothing. We apply the main
effects approach to six composite indicators, including
the HDI and two popular league tables of university performance.
We find that in some cases, a linear model can give a
reasonable estimate of the main effects, but in other
cases the non-parametric fit must be preferred.
Further, we find that
\textit{nominal weights hardly ever coincide with main effects}.
We propose to summarize this deviation in a discrepancy statistic,
which can be used by index developers and users alike to gauge
the gap between the effective and the target importance of each variable.

We also pose the question of whether the target importance
stated by the developers is actually attainable by appropriate
choice of nominal weights; we call this the `inverse problem'.
We find that in most instances the correlation
structure prevents developers from obtaining the stated
importance by changing the nominal weights within the set of
nonnegative numbers with sum equal to 1. These findings may offer a useful
insight to users and critics of an index, and a stimulus to
its developers to try alternative, possibly non-compensatory,
aggregation strategies.

Our proposed measure of importance is also in line with current
practice in Sensitivity Analysis. Recently, some of the present authors
proposed a global sensitivity analysis approach to test the robustness
of a composite indicator, see \citet{Saisana-al05,Saisana-al11}; this approach performs an
error propagation analysis of all sources of uncertainty which can affect
the construction of a composite indicator. This analysis might be called `invasive' in
that it demands all sources of uncertainty to be modeled
explicitly, e.g. by assuming alternative methods to impute missing
values, different weights, different aggregation strategies; the method
may also test the effect of including or excluding individual variables from
the index.

In contrast, the approach suggested in this paper is non-invasive, because it
does not require explicit modeling of uncertainties.
The proposed measure also requires minimal assumptions, in the sense that
it exists whenever second moments exist. Moreover, it takes
the data correlation structure into account.
When this analysis is performed by the developers themselves, it adds to the understanding
-- and ultimately to the quality, of the index. When performed ex-post by a third party
on an already developed index, this procedure may reveal un-noticed features of the composite indicator.

The paper is organized as follows: the rest of Section \ref{sec_Introd} reports a motivating
example and discusses related work. Section \ref{W-vs-I} describes linear composite indicators.
Section \ref{OurProposal} defines the main effects and discusses their estimation.
It also defines a discrepancy statistic
between main effects and nominal weights.  Finally it discusses the inversion of the map
from nominal weights to main effects. Section \ref{Case-Studies-var}
presents detailed results for six indices: the 2009 Human
Development Index (2009 HDI), the Academic Ranking of World
Universities by Shanghais Jiao Tong University (ARWU), the university
ranking by the Times Higher Education Supplement (THES), the 2010
Human Development Index (2010 HDI), the Index of African
Governance (IAG) and the Sustainable Society Index (SSI).
Section \ref{Conclusions} contains a discussion and conclusions.
A solution to the inverse problem is reported in the Appendix.

\subsection{Motivating example}\label{sec1_motiv-exmpl}
In weighted arithmetic averages, nominal weights are
communicated by developers and perceived by users
as a form of judgement of the relative importance of the different variables,
including the case of equal weights where all variables are assumed to be
equally important. When using `budget allocation', a strategy to assign
weights, experts are given a number of tokens, say $100$, and asked to
apportion them to the variables composing the index, assigning more tokens
to more important variables. This is a vivid example
of how weights are perceived and used as measures of importance.
However, the relative importance of variables depends on the
characteristics of their distribution (after normalization) as well as their
correlation structure, as we illustrate with the following example. This
gives rise to a paradox, of weights being perceived by users as
reflecting the importance of a variable, where this perception can
be grossly off the mark.

Consider a University Dean who is asked to evaluate the performance of
faculty members, giving equal importance to indicators of
publications $x_{1}$, of teaching $x_{2}$ and
of office hours and administrative work $x_{3}$.
Hence she considers an equally-weighted index, $y=\frac{1}{3}%
(x_{1}+x_{2}+x_{3})$, and
she employs $R_{i}^{2}:=\mathrm{corr}^{2}(y,x_{i})$
in order to measure the association between the index $y$ and
each of the $x$ variables ex post.

We consider two different situations, which illustrate the
influence of variances and of correlations
of the $x$ variables on the performance of faculty members.
In both situations, we let the variables $x_{1}$, $%
x_{2}$, $x_{3}$ be jointly normally distributed with mean zero.
First assume that the variance of $x_{1}$ is equal to $7$ while $x_{2}$
and $x_{3}$ have unit variances, and that the $x_{j}$ variables are
uncorrelated; the value 7 is chosen here in order to make the
variance of $y$ equal to 1. We then find
\[
R_{1}^{2}=\frac{7}{9}\approx 0.778,\qquad R_{2}^{2}=R_{3}^{2}=\frac{1}{63}%
\approx 0.016,
\]%
which implies that the importance (as measured by $R_{i}^{2}$) of the
variables $x_{2}$ and $x_{3}$ relative to $x_{1}$ is equal to
$1/49\ \approx  0.020$. This shows how variances can greatly affect this measure of
importance. We conclude that the Dean needs to do something about the indicators'
variances before computing the index.

Changing the weights from $1/3$ to
$1/(c\sqrt{\sigma_{ii}})$, where $c:=\sum_{i=1}^3 1/\sqrt{\sigma_{ii}}$
and $\sigma_{ii}$ is the variance of $x_i$ would compensate for
unequal variances; this corresponds to standardizing indicators before
aggregation. In current practice, composite indicators builders prefer to normalise indicators
before aggregation, for instance dividing by the highest score.
Going back to the Dean's example, the yearly number of administration hours
can be divided by the total number of hours within a year, delivering $x_3$ as the \textit{fraction}
of administration hours. We remark that, in general, normalised scores present different variances.

Consider next the situation where $x_{1}$, $x_{2}$, $x_{3}$ are standardized, i.e.
have all unit variances. Assume also that the correlations
$\rho _{ij}:=\mathrm{corr}(x_{i},x_{j})$ are all equal to zero,
except $\rho _{23}=\rho _{32}>0$.
Simple algebra shows that
\[
R_{1}^{2}=\frac{1}{3+2\rho _{23}},\qquad R_{2}^{2}=R_{3}^{2}=%
\frac{(1+\rho _{23})^2}{3+2\rho _{23}},\qquad
\frac{R_{1}^{2}}{R_{2}^{2}}=\frac{1}{(1+\rho _{23})^2},
\]
i.e. that the importance of indicators $x_2$ and $x_3$ is the same;
this is a general property of standardized indicators. Note that
the importance of indicators $x_2$ and $x_3$ is greater that the one of $x_1$, because
$\rho _{23}>0$.
Taking for instance $\rho _{23}=0.7$, one finds
\[
R_{1}^{2}=\frac{5}{22}\approx 0.227,\qquad R_{2}^{2}=R_{3}^{2}=%
\frac{289}{440} \approx 0.657, \qquad \frac{R_{1}^{2}}{R_{2}^{2}}=\frac{100}{289}\approx
0.346.
\]

One may well imagine a faculty member looking at
the relative importance of $x_{1}$ with respect
to $x_{2}$, complaining that research has become dispensable,
because -- although the index's formula seems to suggest that
all variables are equally important -- in fact \textit{teaching is valued
more than publications by a factor of} 3.
In this second situation, even if the Dean has standardized
the variables measuring publications $x_{1}$, teaching $x_{2}$ and administration $x_{3}$,
the last two have a higher influence on the faculty performance indicator $y$
due to their correlation.

\smallskip
This example describes different situations which generate the paradox.
The occurrence of different variances is one such situation; this is a problem also in practice,
because usually individual indicators are normalized to be between
0 and 1 or 0 and 100, and hence they have different variances in general.
Also when correcting for different variances using standardized indicators,
however, the paradox can be generated by correlations. This
is of practical concern as well, because different individual indicators are usually correlated.

The paradox illustrated by the preceding example equally applies when the index's architecture is made
of pillars, each pillar aggregating a subset of variables. An hypothetical
sustainability index could have environmental, economic, social and
institutional pillars, and equal weights for these four pillars would flag
the developers' belief that these dimensions share the same 
importance. Still one of the four pillars with a weighting in principle of $25\%$
could contribute little or nothing to the index, e.g. because the variance
of the pillar is comparatively small and/or the pillar is not correlated to the
remaining three. A case study of this nature is discussed later in the present work.

\subsection{Related work}
\label{Alternatives}The connection of the present paper with global sensitivity analysis has been discussed above.
A related approach to measure variable importance in linear aggregations
is the one of `effective weights', introduced in the psychometric literature
by \citet{StanleyWang1968}, \citet{WangStanley70}. The effective weight of a
variable $x_i$ is defined as the covariance between
$w_i x_i$ and the composite indicator $y=\sum_{i=1}^k w_i x_i$ divided by its
variance, i.e. $\epsilon_i := \mathrm{cov}(y,w_ix_i) /\mathrm{V}(y) $. The same
approach has been employed in recent literature in global sensitivity analysis, see e.g. \citet{LiRabitz2010}.

Effective weights $\epsilon_i$ are, however, not necessarily positive, and hence they make
an improper apportioning of the variance $\mathrm{V}\left( y\right) $: $\epsilon_i$ cannot be interpreted as a `bit'
of variance.
On the contrary, the measure of importance $S_i$ proposed
in this paper (i.e. Pearson's correlation ratio) is
always positive and can be interpreted as the fractional reduction
in the variance of the index that could be achieved (on
average) if variable $x_i$ could be fixed.
$S_i$ also fits into an ANOVA variance decomposition framework,
see \cite{Salt2002} for a discussion.

Moreover, effective weights assume that the dependence structure of
the variables $x_i$ is fully captured by their covariance structure, as
in linear regression. As we show in the following, the relation between
the index and its components may well be nonlinear, and the measure
of importance proposed in this paper extends to this case as well.
The case-studies reported in Section \ref{Case-Studies-var} show that
nonlinearity is often the rule rather than the exception.
In the case of a linear relation between $y$ and $x_i$, our measure
$S_i$ reduces to $R_i^2$, the square of $\mathrm{corr}(y,x_i)$,
used in the example above; hence
in this case, the present approach leads to a simple transformation of the effective weights.

For some indices, such as the Product Market Regulation Index (see
\citet{Nicoletti00}), Principal Component Analysis
(PCA) has been used to select aggregation weights. PCA chooses weights that
maximize (minimize) the variance of the index, and hence weights do not
reflect the normative aspects of the definition of the index. Consequently, weights
are difficult to interpret and to communicate, and as a result the use of PCA in this
context is not widespread.
The same Product Market Regulation Index moved from the use of PCA to a simpler and more transparent technique for linear aggregation after a statistical analysis of the implications of such a change
\cite[]{Nardo2009}.

\section{Weights and importance}
\label{W-vs-I}Consider the case of a composite indicator $y$ calculated as a
weighted arithmetic average of $k$ variables $x_{i}$
\begin{equation}
y_{j}=\sum_{i=1}^{k}{w_{i}}{x_{ji},\qquad }j=1,2,\cdots ,n  \label{def}
\end{equation}%
where $x_{ji}$ is the normalized score of individual $j$ (e.g., country)
based on the value $X_{ji}$ of variable $X_{\cdot i}$, $i=1,\dots ,k$ and $%
w_{i}$ is the nominal weight assigned to variable $X_{\cdot i}$.
The most common approach is to normalize original variables, see \citet{Bandura08}, by the min-max normalisation method
\begin{equation}
{x_{ji}}=\frac{X _{ji}-X _{\min ,i}}{X _{\max ,i}-X_{\min ,i}},
\label{miniMax}
\end{equation}%
where $X_{\max ,i}$ and $X_{\min ,i}$ are the upper and lower values
respectively for the variable ${X_{\cdot i}}$; in this case all scores $%
x_{ji}$ vary in $[0,1]$. Here we indicate the transformation (\ref{miniMax})
as `normalisation';
the normalised variables in (\ref{miniMax}) are
denoted as $x_{\cdot i}$. We let $\mu _{i}:=\mathrm{E}(x_{ji})$ and $\sigma _{ii}=\mathrm{V}(x_{ji})$
indicate their expectation and variance respectively.
In the following, we replace $X_{\cdot i}$ and $x_{\cdot i}$ by $X_{i}$
and $x_{i}$ respectively, unless needed for clarity.

Observe that the normalisation (\ref{miniMax})
implies a fixed scale of the individual indicators; this is useful for instance
for comparability in repeated waves of the same index.
However, normalisation
does not imply any standardization of
different $x_{\cdot i}$\ variables, which hence have
different means $\mu _{i}$ and variances $\sigma _{ii}$ in general.

A popular alternative to the min-max normalization in (\ref{miniMax}) is given by standardization
\begin{equation}
{x_{ji}}=\frac{X_{ji}-\mathrm{E}\left( X_{ji}\right) }{\sqrt{\mathrm{V}\left(
X_{ji}\right) }},  \label{Standardization}
\end{equation}%
where $\mathrm{E}\left( X_{ji}\right) $ and $\mathrm{V}\left(
X_{ji}\right) $ are the mean and variances of the original variables $X_{\cdot i}$. When
standardized, all $x_{i}$ have the same mean and variance, $\mu _{i}=0$,
$\sigma _{ii}=1$ for all $i$, removing one source of heterogeneity among variables.
However, standardization does not affect the correlation
structure of the variables $X_{i}$ (or $x_{i}$). Both transformations (\ref{miniMax}) and
(\ref{Standardization}) are invariant to the choice of unit of measurement
of $X_{i}$, see \cite[Chapter 1]{Hand2009}.

While standardization may appear a better approach
than normalization, statistically, there are advantages and disadvantages of both.
For example standardization may be expected not to work so well
when the distribution is very skewed or long tailed. Moreover it does not enhance comparison across
different waves of the same aggregate indicator over the years, if the mean and variances used in
(\ref{Standardization}) change over time.
Also one cannot achieve \emph{both} standardization and normalization
at the same time through a linear transformation of $X_{i}$. This implies that index developers
suffer the unwanted disadvantages of the chosen transformation.

Whatever the transformation,
in the following we denote the
column vector of scores of unit $j$ as $\mathbf{x}_{j}:=(x_{j1},\dots
,x_{jk})^{\prime }$ and indicate by $\mathbf{\mu }:=(\mu _{1},\dots ,\mu
_{k})^{\prime }$ and $\mathbf{\Sigma }:=(\sigma _{it})_{i,t=1}^{k}$ the
corresponding vector of means and the implied variance-covariance matrix.
The weight, $w_{i}$, attached to each variable, $x_{i}$,
in the aggregate is meant to appreciate
the importance of that variable with respect to the concept being measured.
The vector of weights $\mathbf{w}:=(w_{1},\dots ,w_{k})^{\prime }$ is
selected by developers on the basis of different strategies, be those
statistical, such as PCA, or
based on expert evaluation, such as analytic hierarchy process, see \citet{Saaty80,Saaty87}.

In what follows we indicate by $\zeta _{i\ell}^{2}$ the target relative importance of
indicators $i$ and $\ell$. When this is not explicitly stated,
the ratios $w_{i}/w_{\ell}$ can be taken to be the
`revealed target relative importance'.
In fact $w_{i}/w_{\ell}$ is a measure of
the substitution effect between $x_{i}$ and $x_{\ell}$, i.e. how much $x_{\ell}$
must be increased to
offset or balance a unit decrease in $x_{i}$, see \citet{Decancq-Lugo2010}.
For simplicity of notation and without loss of generality,
we assume that the maximal weight is assigned to indicator 1,
i.e. that $w_{1}\geq w_{i}$ for $i=2,\dots,k$,
and we consider $\zeta_{i}^{2}:=\zeta _{i1}^{2}$.

Note that the previous discussion applies to pillars as well as to
individual variables, where a pillar is defined as an aggregated subset of
variables, identified by the developers as representing a salient --
possibly latent, or normative -- dimension of the composite indicator.%

\section{Measuring importance}\label{OurProposal}
\subsection{Measures of importance}
In this paper we propose a variance-based measure of
importance. We note that
\begin{equation}
\mathrm{E}\left( y\right) =\mathbf{w}^{\prime }\mathbf{\mu },\qquad \mathrm{V%
}\left( y\right) =\mathbf{w}^{\prime }\mathbf{\Sigma w,}
\label{eq_meanY_varY}
\end{equation}%
where, if (\ref{Standardization}) is used, $\mathrm{E}\left( y\right) =0$
and the diagonal elements of $\mathbf{\Sigma }$ are equal to 1; here we have
dropped the subscript $j$ in $y_{j}$ for conciseness. In the following, we
focus attention on the variance term.

Following \citet{Pearson05}, we consider the question
`what would be the average variance of $y$, if
variable $x_{i}$ were held fixed?' This question leads to consider
\[
\mathrm{E}_{x_{i}}\left( \mathrm{V}_{\mathbf{x}_{\sim i}}\left( y\mid
x_{i}\right) \right) ,
\]
where $\mathbf{x}_{\sim i}$ is defined as the vector containing
all the variables in $\mathbf{x}$ except variable $x_{i}$.
Owing to the well known identity
\[
\mathrm{V}_{x_{i}}\left( \mathrm{E}_{\mathbf{x}_{\sim i}}\left( y\mid
x_{i}\right) \right) +\mathrm{E}_{x_{i}}\left( \mathrm{V}_{\mathbf{x}_{\sim
i}}\left( y\mid x_{i}\right) \right) =\mathrm{V}\left( y\right)
\]
we can define the ratio of $\mathrm{V}_{x_{i}}\left( \mathrm{E}_{\mathbf{x}%
_{\sim i}}\left( y\mid x_{i}\right) \right) $ to $\mathrm{V}\left( y\right) $
as a measure of the\ relative reduction in variance of the composite
indicator to be expected by fixing a variable, i.e.
\begin{equation}
S_{i}\equiv \eta _{i}^{2}:=\frac{\mathrm{V}_{x_{i}}\left( \mathrm{E}_{\mathbf{%
x}_{\sim i}}\left( y\mid x_{i}\right) \right) }{\mathrm{V}\left( y\right) }.
\label{SiBis}
\end{equation}
The notation $S_{i}$ reflects the use of this measure as a first order
sensitivity measure (also termed `main effect') in sensitivity analysis, see
\citet{SaltTara02}. The notation $\eta _{i}^{2}$ reflects
the original notation used in \citet{Pearson05}; he called it `correlation ratio $\eta ^{2}$'.

The conditional expectation $\mathrm{E}_{\mathbf{x}_{\sim i}}\left( y\mid
x_{i}\right) $ in the numerator of (\ref{SiBis}) can be any nonlinear
function of $x_{i}$; in fact $f_{i}\left( x_{i}\right) :=%
\mathrm{E}_{\mathbf{x}_{\sim i}}\left( y\mid x_{i}\right)
=w_{i}x_{i}+\sum_{\ell=1,\ell\neq i}^{k}w_{\ell}\mathrm{E}_{\mathbf{x}_{\sim
i}}\left( x_{\ell}\mid x_{i}\right) $, where the latter conditional
expectations may be linear or nonlinear in $x_{i}$. For
the connection of $f_{i}\left( x_{i}\right) $ to global sensitivity analysis see \citet{SABook2008}.

In the special case of $f_{i}\left( x_{i}\right) $ linear in $x_{i}$,
we find that $S_{i}$
reduces to $R_{i}^{2}$, where $R_{i}$ is the product-moment
correlation coefficient of the regression
of $y$ on $x_{i}$. In fact, it is well known that when
$f_{i}$ is linear, i.e. $f_{i}\left(
x_{i}\right) =\alpha _{i}+\beta _{i}x_{i}$, it coincides with
the $L_2$ projection of $y$ on $x_i$, which implies that
$\beta _{i} = \mathrm{cov}(y,x_i) / \sigma_{ii}$, see e.g. \citet{Wooldridge:10}.
Hence $S_{i}$ has the form
$S_{i}=\mathrm{V}_{x_{i}}\left( \beta _{i}x_{i}+\alpha _{i}\right) /\mathrm{V}( y )$
and one finds
$S_{i}=\beta _{i}^{2}\sigma _{ii}/\mathrm{V}( y ) =
\mathrm{cov}^2(y,x_i)/ ( \sigma _{ii} \mathrm{V}( y )) =R^2_i$.

A further special case corresponds to $f_{i}$ linear and
$\mathbf{x}$ made of uncorrelated components.
We find
$\mathrm{cov}(y,x_i) = \sum_{t=1}^{k}w_{t}\sigma_{ti}$
and $\mathrm{V}(y) = \sum_{t=1}^{k}w^2_{t}\sigma_{tt}$
so $S_{i}=w_{i}^{2}\sigma
_{ii}/\sum_{t=1}^{k}w_{t}^{2}\sigma _{tt}$.
The main difference between the uncorrelated and the correlated case is that
in the former $\sum_{i=1}^{k}S_{i}=1$ because $S_{i}=w_{i}^{2}\sigma
_{ii}/\sum_{h=1}^{k}w_{h}^{2}\sigma _{hh}$, while for the latter $%
\sum_{i=1}^{k}S_{i}$ might well exceed one, see e.g. \citet{SaltTara02}.
We note that in general $S_{i}$ can still be high
also when $R_{i}^{2}$ is low, e.g. in case of a
non-monotonic U-shaped relationship for $f_{i}( x_{i} )  $. Hence
in general $f_{i}( x_{i} ) $ needs to be estimated
in a nonparametric way, see Section \ref{sec_estimation}.

As these special cases illustrate, $S_{i}$ is quadratic measure in terms of the weights $w_{j}$
for linear aggregation schemes (\ref{def}); this follows
from its definition as a variance-based measure.
The main effect $S_{i}$ is an appealing measure of importance
of a variable (be it indicator or pillar) for several reasons:
\begin{itemize}
\item it offers a precise definition of importance of a variable, that is
`the expected fractional reduction in variance of the composite indicator
that would be obtained if that variable could be fixed';
\item it can be applied when relationships between
the index and its components
are linear or nonlinear. Such nonlinearity may be the
effect of nonlinear aggregation
(e.g. Condorcet-like, see \citet{Munda2008}) and/or of nonlinear
relationships among the single variables. It can be used
regardless of the degree of correlation between variables.
Unlike the Pearson or Spearman correlation coefficients,
it is not constrained by assumptions of linearity or monotonicity;
\item it is not invasive, that is no changes are made to the composite
indicator or to the correlation structure of the indicators,
unlike e.g. the error propagation analysis presented in \citet{Saisana-al05}.
While the error propagation can be considered as a stress test of the index,
the present approach is a test of its internal coherence.
\end{itemize}

\subsection{Estimating main effects}\label{sec_estimation}
In this subsection we consider estimating the
main effects and focus on the 2009 HDI to illustrate our approach. In
Section \ref{Case-Studies-var} we describe the six case-studies of our approach in detail.

\begin{figure}
\centering
\makebox{\includegraphics[scale=0.60]{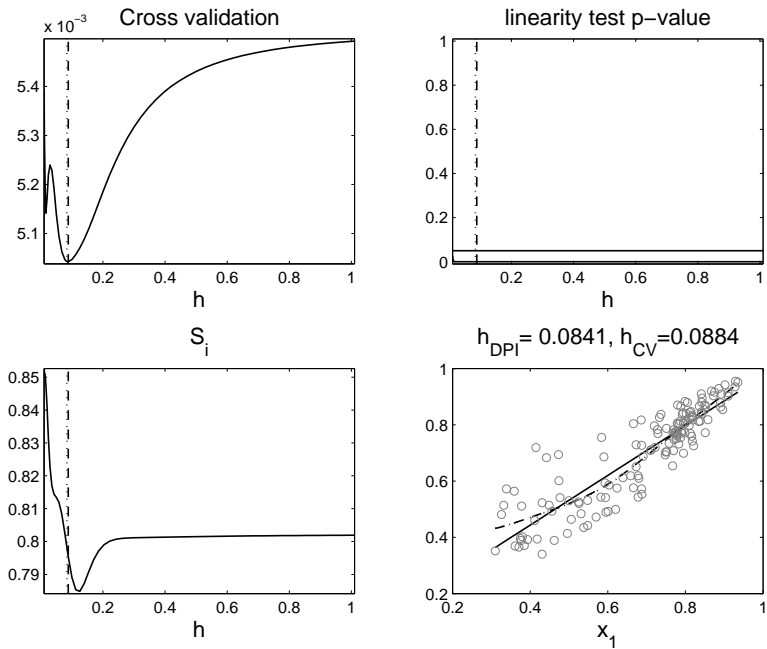}}
\caption{\label{Fig_HDI9-1} 2009 HDI ($y$), Life expectancy ($x_1$).
Upper left: Cross validation criterion as a function of the smoothing
parameter $h$; upper right: linearity test $p$-value as a function of $h$;
lower left: main effects $S_i$ as a function of $h$; lower right: cross plot
of $y$ versus $x_1$ with linear fit and local linear fits for $h_{DPI}$
(direct plug-in, dotted line) and $h_{CV}$ (cross validation, dashed line).
The values of $h_{DPI}$ and $h_{CV}$ are plotted as vertical lines
in the first 3 panels (dotted lines and dashed lines respectively).}
\end{figure}

In sensitivity analysis, the estimation of $S_{i}$ is an active research field. $S_{i}$ can be
estimated from design points: \citet{Sobol1990-1993,Salt2002,Saltetal_2010};
Fourier analysis: \citet{Tarantola-al06,Plischke10,Xuetal2011}, or others. Many
nonparametric estimators can be used to estimate $f_{i}( x_{i}) $, such as
State Dependent Regression: \citet{Ratto-al07,RattoPa10}.

In the present work we employ a nonparametric, local-linear, kernel regression to
estimate $m(\cdot ):=f_{i}( \cdot) $, and then use it in (\ref{SiBis}) to estimate $S_i$, replacing
the variances in the numerator and denominator with the corresponding sample
variances, i.e. using
$\sum_{j=1}^n ( m_{j}-\bar{m})^2 /\sum_{j=1}^n (y_j - \bar{y})^2$, where
$\bar{y}:=n^{-1}\sum_{j=1}^n y_j$, $\bar{m}:=n^{-1}\sum_{j=1}^n m_j$,
$m_j := \hat{m}(x_{ji})$ and $\hat{m}(\cdot)$ is the estimate of $m(\cdot ):=f_{i}( \cdot) $.

Local linear kernel estimators achieve automatic boundary
corrections and enjoy some typical optimal properties, that are superior to
Nadaraya-Watson kernel estimators, see \citet{RuppertWand:1994} and
reference therein. As a result, local linear kernel smoother are often considered
the standard nonparametric regression method, see e.g. \citet{BowmanAzzalini:97}.

The local linear nonparametric kernel regression is indexed by a bandwidth parameter
$h$, which is usually held constant across the range of value for $x_i$. For large $h$,
the local linear nonparametric kernel regression converges to the linear least squares
fit. This allows us to interpret $1/h$ as the deviation from linearity;
it suggests that we investigate the sensitivity of the estimation of $S_i$ to
variation in the bandwidth parameter $h$. In order to make this dependence explicit
we write $S_i(h)$ to indicate the value of $S_i$ obtained by a local-linear kernel regression
with bandwidth parameter $h$. In the application we use a Gaussian kernel.

The choice of the smoothing parameter $h$ can be based either on cross-validation (CV) principles (see \citet{BowmanAzzalini:97}) or on plug-in choices for the smoothing parameter,
such as the ones proposed in \citet{RuppertSW:2001}. We describe these approaches in turn,
starting with cross validation.
Let $\hat{m}(x)$ indicate the local linear nonparametric kernel estimate for $f_{i}( x) $ at $x_i = x$
based on all $n$ observations, and let
$\hat{m}_{-j}(x)$ be the same applied to all data points except for the one with index $j$; then
the least-squares Cross Validation criterion for variable $x_i$ is defined as
$$
CV(h) = \frac{1}{n}\sum_{j=1}^n \left( y_{j} -\hat{m}_{-j}(x_{ji})\right)^{2}
$$

The optimal value for the CV criterion is given by the bandwidth $h_{CV}$ corresponding
to the minimum of $CV(h)$. In practice, a grid $\mathcal{H}$ of possible values for $h$ is
considered, and the minimum of the function $CV(h)$ is found numerically.
In the application we chose the grid of $h$ values as follows: we defined a regular grid of
50 values for $u:=\sqrt{h}$ in the range from 0.1 to 5. The values for $h$ were then
obtained as $h=a+u^2/b$, for index-specific constants $a$ and $b$; the resulting set
of values in this grid is denoted $\mathcal{H}$ in what follows.

The default values
for indices with range from 0 to 10 or 100 were $a=0.05$, $b=1$,
so that $.06 < h \leq 25.05$; for indices with range from 0 to 1, (namely 2009 and 2010 HDI ), we chose $a=0.01$, $b=25$, so that $.01 < h \leq 1.01$.
In some cases CV$(h)$ attains its minimum at the right end of the grid $\mathcal{H}$;
This happened both for ARWU$\{1,2,3\}$ and THES$\{4,6\}$, see Table \ref{Table_Si_var-h}, as well as
for IAG$\{2,5\}$ and SSI$\{2\}$, see Table \ref{Table_Si_pil-hi}, where the digits in braces refer
to the subscript $i$ of the $x_i$ variables.
In these cases, in practice, a linear regression fit would not be worse than the fit
of the local linear kernel estimator, according to the CV criterion.

In the implementation of the CV criterion, when a local linear kernel regression implied
a row of the smoothing matrix with numerical `divisions by zero',
we replaced it with a local mean (Nadaraya-Watson) estimator. When also
the latter would imply numerical divisions by zero, we replaced the row of the smoothing matrix with a
sample leave-one-out mean.

\begin{figure}
\centering
\makebox{\includegraphics[scale=0.60]{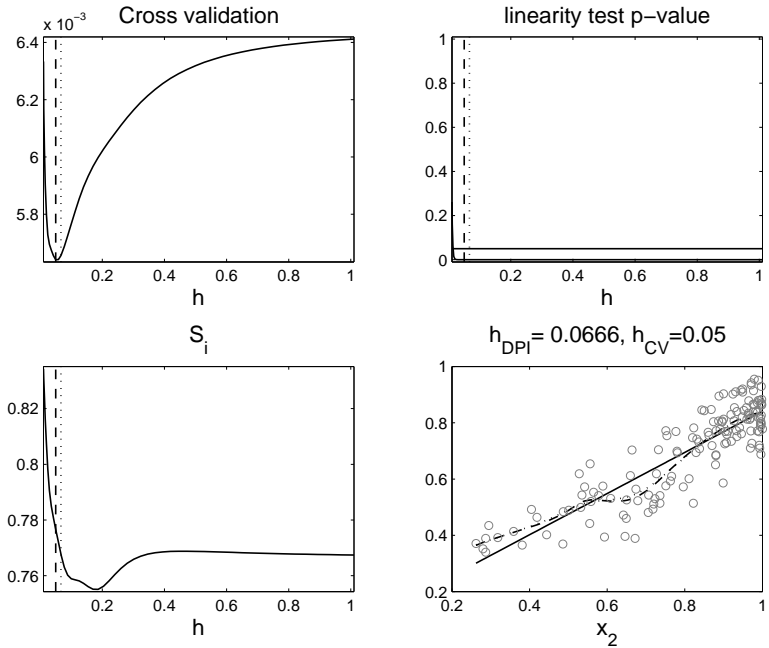}}
\caption{2009 HDI ($y$), Adult literacy ($x_2$). See caption of Fig \ref{Fig_HDI9-1}.}
\label{Fig_HDI9-2}
\end{figure}

An alternative choice of bandwidth is given by plug-in-rules. One popular choice is given
by the `direct plug-in' selector $h_{DPI}$ introduced by \citet{RuppertSW:2001}, which
minimizes the Asymptotic Mean Integrated Squared Error for the local linear Gaussian
kernel smoother, on the basis of the following preliminary estimators. Let
$\theta_{rs}:=\mathrm{E}(m^{(s)}m^{(r)})$, where $m^{(r)}(x)$ is the $r$-th
derivative of $m(x)$. The range of $x_i$ is partitioned into $N$ blocks and a
quartic is fitted on each block. Using this estimation, an estimate for $\theta_{24}$
is found, along with an estimator for the error variance $\sigma^2 := \mathrm{E}(y_j-m(x_{ji}))^2$.
These estimates are then used to obtain a plug-in bandwidth $g$, which is used
in a local cubic fit to estimate $\theta_{22}$ and to obtain a different
plug-in bandwidth $\lambda$. The $\lambda$ bandwidth is then used in a final
local linear kernel smoother to estimate $\sigma^2$, which is fed into the final formula
for $h_{DPI}$, along with the previous estimate of $\theta_{22}$. The choice of $N$,
the number of blocks, is obtained minimizing Mallow's $C_p$ criterion
over the set $\{1,2,\dots,N_{\max}\}$, where $N_{\max} = \max\{\min( \lfloor n / 20\rfloor, N^{\ast}),1\}$.

In the application we chose $N^{\ast} = 5$ as suggested by \citet{RuppertSW:2001};
in case of numerical instabilities, we decreased $N^{\ast}$ to 4. Moreover we
performed an $\alpha$-trimming in the estimation of $\theta_{24}$ and
$\theta_{22}$ with $\alpha=0.05$.
Because the choice of bandwidth can be affected by values at the end of the
$x$-range, we only considered pairs of observations
for which $x>0$ in the choice of bandwidth, both for the $CV$ criterion and the $DPI$ criterion.

\begin{figure}
\centering
\makebox{\includegraphics[scale=0.60]{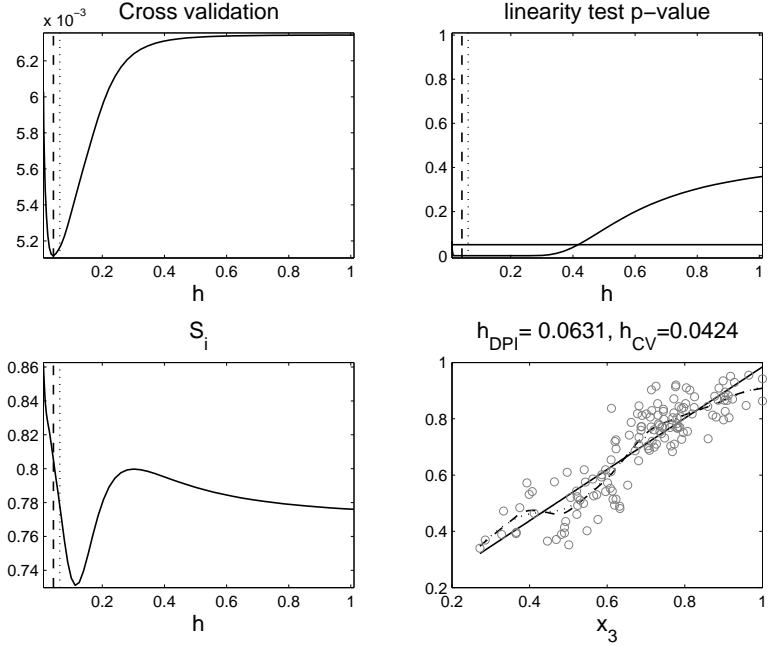}}
\caption{2009 HDI($y$), Enrolment in education ($x_3$). See caption of Fig \ref{Fig_HDI9-1}.}
\label{Fig_HDI9-3}
\end{figure}

\begin{figure}
\centering
\makebox{\includegraphics[scale=0.60]{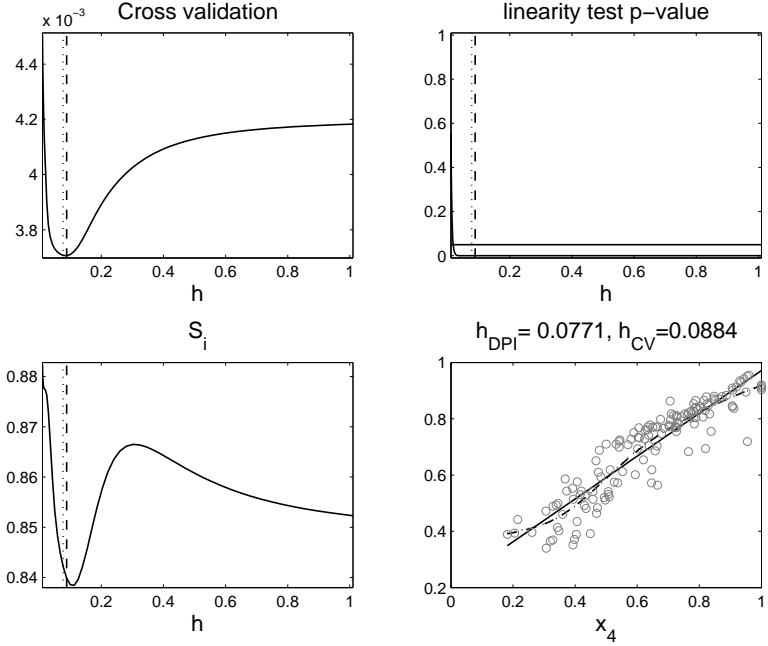}}
\caption{2009 HDI ($y$), GDP per capita ($x_4$). See caption of Fig \ref{Fig_HDI9-1}.}
\label{Fig_HDI9-4}
\end{figure}

The resulting choice of bandwidth
$h_{DPI}$ was sometimes very close to $h_{CV}$, as in the case for the
2009 HDI, which is depicted in Fig.~\ref{Fig_HDI9-1}-\ref{Fig_HDI9-4},
where each figure refers to one of the four $x_i$ indicators used in the
construction of the 2009 HDI.
Fig.~\ref{Fig_HDI9-1} refers to  the $x_1$ indicator (life expectancy),
and contains four panels, which report --
counterclockwise from upper-right --
the $p$-value of the linearity test introduced below,
the cross validation criterion $CV$,
the $S_1$ measure and the regression cross plot.
The first 3 graphs show functions of the bandwidth parameter $h$,
while the final one has the values of $x_1$ on the
horizonal axis.
Fig.~\ref{Fig_HDI9-2}-\ref{Fig_HDI9-4} have the same format, and refer to
indicators $x_2, x_3$ and $x_4$.

Tables~\ref{Table_Si_var-h} and \ref{Table_Si_pil-hi} report the
selected values of $h_{CV}$ and $h_{DPI}$ for the 2009 HDI
and for the other 5 indices, described in detail in Section \ref{Case-Studies-var}.
It can be seen that the values of $h_{CV}$ sometimes
differed from $h_{DPI}$ by several orders of magnitude.

As in many other contexts, in the estimation of main effects $S_i$ the
linear case is a relevant reference model, and one would like to address
inference on $S_i$ and on the possible linearity of $f_i(x_i)$ jointly.
To this end we implemented the test for linearity proposed in
\citet[Chapter 5]{BowmanAzzalini:97}. The fit of the linear kernel
smoother can be represented as
$\hat{\mathbf{y}} = \mathbf{S} \mathbf{y}$, where the matrix $\mathbf{S}$
depends on all values $x_{ji}$, $j=1,\dots,n$. A test of linearity can be based
on the $F$ statistic,  $F:=(RSS_0-RSS_ 1)/RSS_1$, that compares the residual
sum of squares under the linearity assumption $RSS_0$ with the one
corresponding to the local linear kernel smoother $RSS_1$.
Letting $F_{\mathrm{obs}}$ indicate the value of the statistic,
the $p$-value of the test is computed as the probability that
$\mathbf{z}^{\prime} \mathbf{C}\mathbf{z} >0$ where $\mathbf{z}$
is a vector of independent standard Gaussian random variables and
$\mathbf{C}:=\mathbf{M}(\mathbf{I}-(1+F_{\mathrm{obs}})\mathbf{A})\mathbf{M}$
with $\mathbf{A} = (\mathbf{I}-\mathbf{S})^{\prime}(\mathbf{I}-\mathbf{S})$,
$\mathbf{M} = \mathbf{I}-\mathbf{X}(\mathbf{X}^{\prime}
\mathbf{X}) ^{-1}
\mathbf{X}^{\prime}$ and $\mathbf{X}$ equal to
the linear regression design matrix, with first column equal to the constant
vector and the second column equal to the values of $x_{ji}$, $j=1,\dots,n$.

\citet{BowmanAzzalini:97} suggest approximating the quantiles of the quadratic
form with the distribution of $a \chi^2_b + c$, where $a$, $b$ and $c$ are obtained
by matching moments of the quadratic form and the $a \chi^2_b + c$ distribution;
here $\chi^2_b$ represents a $\chi^2$
distribution with $b$ degrees of freedom.
We implemented this approximation; the upper right panels in
Fig.~\ref{Fig_HDI9-1}-\ref{Fig_HDI9-4}
report the resulting $p$-values of the test as a function of $h$ for the 2009 HDI.
It can be seen for some $x_i$ variable the test rejects the linearity hypothesis
for all values of $h$ in the grid $\mathcal{H}$, and for
some other pairs the test rejects only for a subset of
$\mathcal{H}$. In a few other pairs, the test never rejects
for all $h \in \mathcal{H}$. Results for the linearity test are
reported in Tables~\ref{Table_Si_var-h} and \ref{Table_Si_pil-hi}
for selected values of $h$,
both the 2009 HDI and for 5 other indices,
described in detail in Section \ref{Case-Studies-var}.

To show sensitivity of the main effects $S_i$ to the
smoothing parameter $h$, we also computed the $S_i(h)$ index
as a function of $h$. We also recorded the min and max values
obtained for $S_i(h)$ varying $h$ in $\mathcal{H}$; we denote
these values as $S_{i,\min}$, $S_{i,\max}$. We report the plot
of $S_i(h)$ as a function of $h$ in the lower left panels of
Fig.~\ref{Fig_HDI9-1}-\ref{Fig_HDI9-4}.

\subsection{Comparing weights and main effects}
\label{sec_compare}In this section we compare revealed or
target relative importance measures $\zeta _{i}^{2}$ with the relative main
effects $S_{i}/S_{1}$.
First notice that, in the independent case, $S_{i}=w_{i}^{2}\sigma
_{ii}/\sum_{h=1}^{k}w_{h}^{2}\sigma _{hh}$, so that $S_{i}/S_{1}=w_{i}^{2}%
\sigma _{ii}/w_{1}^{2}\sigma _{11}$.
When the $x_i$ variables are standardized,
all  $\sigma _{ii}=1$ and hence
$S_{i}/S_{1}=w_{i}^{2}/w_{1}^{2}$.
The relative main
effects $S_{i}/S_{1}$
do not reduce to $\zeta
_{i}^{2}=w_{i}/w_{1}$, except in the homoskedastic case ($\sigma
_{ii}=\sigma _{11}$) when the nominal weights are equal, ($w_{i}=w_{1}$), so
that $w_{i}^{2}\sigma _{ii}/w_{1}^{2}\sigma
_{11}=w_{i}^{2}/w_{1}^{2}=1=w_{i}/w_{1} $. In the general case, $S_{i}$
depends on $\mathbf{w}\ $and $\mathbf{\Sigma }$ in a more complicated way,
and hence there is no reason, a priori, to expect $S_{i}/S_{1}$ to coincide
with $\zeta _{i}^{2}$.

One can compare how the effective relative importance $S_{i}/S_{1}$
deviates from the (revealed) target relative importance $\zeta _{i}^{2}$; to this end we define the
maximal discrepancy statistic $d_m$ as
\begin{equation}
d_m= \max_{i \in \{2,\dots,k\}}{\ \left\vert \zeta _{i}^{2} -\frac{S_{i}}{%
S_{1}}\right\vert }, \label{qmax-equation}
\end{equation}%
In the case of revealed target relative importance, recall that $w_{1}$ is assumed to be the
highest nominal weight $w_{\max}$. In the case when more
than one variable has maximum weight equal to $ w_{\max}$, we selected as reference
variable the one with maximum value for $S_\ell(h_{\ell,DPI})$ with $\ell \in \{1,\dots,k\}$,
i.e. $\ell = \mathrm{argmax}_{ i \in \{1,\dots,k\}}S_i(h_{i,DPI})$ where $h_{i,DPI}$ is the
DPI bandwidth choice for indicator $i$.

The higher the value of $d_m$, the more discrepancy there is between relative target importance
and the corresponding relative main effects.
In $d_m$ we have chosen to
capture the discrepancy by focusing on the maximal deviation; alternatively one can consider
any absolute power mean, $f$-divergence function or distance between the
(un-normalized) distributions $\{\zeta^2_{i}\}$ and $\{S_{i}/S_{1}\}$.
For simplicity, in the following we indicate these distributions
used in the comparison as $\{\zeta^2_{i}\}$ and $\{S_{i}\}$.

Because $d_m$ depends on the choice of bandwidth parameters $h$ in the estimation
of $S_i(h)$, $i=1,\dots,k$ we also calculated bounds on the variation of $d_m$
obtained by varying $h$. Specifically, we computed $d_m$ comparing $\{\zeta^2_{i}\}$ with
$\{S_{i,\ell_i}\}$ choosing $\ell_i$ as either equal to $\min$ or $\max$,
considering all possible combinations. For instance, with $k=2$,
we considered $\{S_{1,\min},S_{2,\min}\}$, $\{S_{1,\min},S_{2,\max}\}$,
$\{S_{1,\max},S_{2,\min}\}$ and $\{S_{1,\max},S_{2,\max}\}$.
Within the distribution of values of $d_m$
obtained in this way, we recorded the minimum and the maximum, denoted as $d_{m,\min}$
$d_{m,\max}$. Table \ref{table_qmax} reports the $d_m$ for $h$ equal
to $h_{DPI}, h_{CV}$ and in the linear case, along with the values $d_{m,\min}$,
$d_{m,\max}$, which provide a measure of sensitivity of $d_m$
with respect to the choice of bandwidth $h$.

To compare the $S_i$ values with the weights $w_i$ graphically, in
Fig \ref{fig-qindex} we re-scale the $S_i$ values to have sum equal to one,
considering $S_i^{\ast} := S_i / c$ with $c:=\sum _{t=1}^k S_t$, which we call `normalized $S_i$'.
In order to visualise bounds for $S_i^{\ast}$, we plot bars with endpoints equal to
$S_{i,\min} / c$ and
$S_{i,\max} / c$; these bars inform on the sensitivity of
$S_i^{\ast}$ with respect to the variation of the bandwidth parameter $h$.

\subsection{Reverse-engineering the weights}\label{Correcting}
This section discusses when it is possible to find nominal weights $w_i$
that imply pre-determined, given values $z_i^2$ for the relative main effects $S_i/S_1$;
here we indicate the target relative importance $z_i^2$ to differentiate it from
$\zeta_i^2$ of the previous sections. This reverse-engineering exercise can
help developers of composite indicators to
anticipate criticism by enquiring if the stated relative importance of
pillars or indicators is actually attainable.

For the purpose of this inversion, we consider the case of $f_{i}(x_{i})$ linear in $x_{i}$;
in this case $S_{i}$ coincides with $R_{i}^{2}$, the
square of Pearson's product moment correlation coefficients between $y$ and $x_{i}$.
The linear case can be seen as a first order approximation to the nonlinear general case;
this choice is motivated by the fact that one can find an exact solution to the
inversion problem of the map from  $w_i$ to $R_{i}^{2}/R_{1}^{2}$ when one allows
weights $w_i$ also to be negative. One expects that the reverse-engineering formula in
the linear case to be indicative of the one based on
a non-linear approach, where the latter would be computationally more demanding.

We wish to find a value $\mathbf{w}^{\ast }:=(w_{1}^{\ast },\dots ,w_{k}^{\ast
})^{\prime }$ for the vector of nominal weights $\mathbf{w}:=(w_{1},\dots
,w_{k})^{\prime }$ such that $R_{i}^{2}/R_{1}^{2}$ equals pre-selected
target values $z_{i}^{2}$, for $i=1,\dots ,k$. We call this the
`inverse problem'. The weights $w_{i}^{\ast }$ are chosen to sum to 1, but
they are allowed also to be negative;\ this choice makes the inverse problem
solvable, and in the Appendix we show that it has a
unique solution, given by
\begin{equation}
\mathbf{w}^{\ast }\mathbf{=}\frac{1}{\mathbf{1}^{\prime }\mathbf{\Sigma }%
^{-1}\mathbf{g}}\mathbf{\Sigma }^{-1}\mathbf{g}.
\label{eq_w_inverse_problem}
\end{equation}%
where $\mathbf{g}$ is a vector with $i$-th entry equal to $g_{i}:=z_{i}\sqrt{%
\sigma _{ii}/\sigma _{11}}$ and $\mathbf{1}$ is a $k$-vector of ones.

Because the solution to this inverse problem is unique, if some of the
weights $w_{j}^{\ast }$ in (\ref{eq_w_inverse_problem}) are negative, it
means that a solution to the inverse problem with all
positive weights does not exist, and hence the targets $z_{i}^{2}$ are not attainable, owing
to the data covariance structure. This can help designers to
re-formulate their targets to make them attainable, and the
stakeholders involved in the use of the composite indicator to evaluate wether 
the individual indicators can have the stated
importance by an appropriate choice of weights.

\begin{table}
\caption{\label{Table_Si_var-h}Bandwidth choice at indicator level.
Bandwidth $h$: $h_{i,CV}$ (cross validation), $h_{i,DPI}$ (direct plug-in, DPI);
$p$-values for the linearity test: $p_{i,CV}$ ($p$-value for $h_{i,CV}$),
$p_{i,DPI}$ ($p$-value for $h_{i,DPI}$); $n$ is the number of observations
with $x_{ji}>0$ used for CV and DPI; $\ast$: right end of the grid $\mathcal{H}$.}
\centering
\fbox{%
\begin{tabular}{llllll}
& {$h_{CV}$} & $p_{CV}$ & $h_{DPI}$ & $p_{DPI}$ & $n$ \\ \hline
\multicolumn{6}{c}{2008 ARWU} \\ \hline
Alumni winning Nobel          & ${25.05}^{\ast}$    & 0.88    & 3.43    & 0.71    & 198    \\
Staff winning Nobel           & ${25.05}^{\ast}$    & 0.59    & 3.13    & 0.27    & 135    \\
Highly cited res.             & ${25.05}^{\ast}$    & 0.00    & 1.15    & 0.00    & 424    \\
Art. in Nature and Science    &  9.05    & 0.00    & 1.78    & 0.00    & 494    \\
Art. in Science and Social CI &  2.94    & 0.00    & 2.26    & 0.00    & 503    \\
Academic perf. (size adj)     &  1.74    & 0.00    & 2.12    & 0.00    & 503    \\\hline
\multicolumn{6}{c}{2008 THES} \\ \hline
Academic review               &  4.46    & 0.00    & 1.74    & 0.00    & 400    \\
Recruiter review              &  5.81    & 0.00    & 2.62    & 0.00    & 400    \\
Teacher/Student ratio         &  4.46    & 0.07    & 4.76    & 0.08    & 399    \\
Citations per faculty         & ${25.05}^{\ast}$    & 0.04    & 2.44    & 0.20    & 400    \\
International staff           &  6.81    & 0.04    & 2.97    & 0.22    & 398    \\
International students        & ${25.05}^{\ast}$    & 0.18    & 4.13    & 0.65    & 399    \\\hline
\multicolumn{6}{c}{2009 HDI} \\ \hline
Life expectancy               &  0.09    & 0.00    & 0.08    & 0.00    & 142    \\
Adult literacy                &  0.05    & 0.00    & 0.07    & 0.00    & 142    \\
Enrolment in education        &  0.04    & 0.00    & 0.06    & 0.00    & 142    \\
GDP per capita                &  0.09    & 0.00    & 0.08    & 0.00    & 142       %
\end{tabular}}
\end{table}

\section{Case studies}\label{Case-Studies-var}
In this section we
apply the statistical analysis that was described in Section \ref{OurProposal} to the six composite
indicators. In Section \ref{sec_sub_casestudies_indicator} we consider the three indices for which
aggregation was performed at indicator level and in Section
\ref{Case-Studies-pillar} we consider the three indices for which aggregation was performed at the
pillar level.

\subsection{Importance at the indicator level}
\label{sec_sub_casestudies_indicator}
We consider the Human Development Index
(HDI) and two well known composite indicators of university performance:
the Academic Ranking of World Universities by Shanghai's Jiao Tong
University (ARWU) and the one associated to the UK's Times Higher Education
Supplement (THES).

\begin{table}
    \caption{\label{Table_Si_var-Si}Main effects at indicator level.
    Nominal weights $w_i$;
    main effects $S_i$:
    $S_{i,lin}:=S_{i} (\infty)$ (linear fit),
    $S_{i,CV}:=S_{i} (h_{CV})$ (cross validation),
    $S_{i,DPI}:=S_{i} (h_{DPI})$ (direct plug-in),
    $S_{i,\min}:=\min_{h\in \mathcal{H}}S_{i} (h)$,
    $S_{i,\max}:=\max_{h\in \mathcal{H}}S_{i} (h)$.}
 \centering
\fbox{%
\begin{tabular}{lllllll}
\hline
& $w_{i}$ & $S_{i,lin}$ & $S_{i,CV}$ & $S_{i,DPI}$ & $S_{i,\min }$ & $%
S_{i,\max }$ \\ \hline
\multicolumn{7}{c}{2008 ARWU} \\ \hline
Alumni winning Nobel          & 0.10    & 0.64    & 0.65    & 0.67    & 0.65    & 0.76    \\
Staff winning Nobel           & 0.20    & 0.72    & 0.72    & 0.73    & 0.72    & 0.80    \\
Highly cited res.             & 0.20    & 0.81    & 0.85    & 0.87    & 0.85    & 0.90    \\
Art. in Nature and Science    & 0.20    & 0.87    & 0.88    & 0.88    & 0.88    & 0.94    \\
Art. in Science and Social CI & 0.20    & 0.63    & 0.70    & 0.70    & 0.64    & 0.90    \\
Academic perf. (size adj)     & 0.10    & 0.71    & 0.76    & 0.75    & 0.72    & 0.88    \\
\hline
\multicolumn{7}{c}{2008 THES} \\ \hline
Academic review               & 0.40    & 0.77    & 0.81    & 0.82    & 0.78    & 0.85    \\
Recruiter review              & 0.10    & 0.45    & 0.54    & 0.54    & 0.46    & 0.62    \\
Teacher/Student ratio         & 0.20    & 0.19    & 0.21    & 0.20    & 0.18    & 0.42    \\
Citations per faculty         & 0.20    & 0.38    & 0.38    & 0.41    & 0.38    & 0.50    \\
International staff           & 0.05    & 0.10    & 0.12    & 0.12    & 0.10    & 0.31    \\
International students        & 0.05    & 0.16    & 0.16    & 0.17    & 0.16    & 0.34    \\
\hline
\multicolumn{7}{c}{2009 HDI} \\ \hline
Life expectancy               & 0.33    & 0.80    & 0.80    & 0.80    & 0.78    & 0.85    \\
Adult literacy                & 0.22    & 0.77    & 0.78    & 0.77    & 0.76    & 0.83    \\
Enrolment in education        & 0.11    & 0.77    & 0.81    & 0.78    & 0.73    & 0.86    \\
GDP per capita                & 0.33    & 0.85    & 0.84    & 0.84    & 0.84    & 0.88   %
\end{tabular}}
\end{table}

\subsection*{University Ranking}
The ARWU, \citet{ARWU2008}, summarizes quality of education, quality of
faculty, research output and academic performance of world universities
using six indicators: the number of alumni of an institution having won Nobel
Prizes or Fields Medals (weight of 10\%), the number of Nobel or Fields
laureates among the staff of an institution (weight of 20\%), the number of
highly cited researchers (weight of 20\%), the number of articles published in
Nature or Science, Science Citation Index Expanded and Social Sciences
Citation Index (weight of 40\%), and finally the academic performance measured
as the weighted average of the above five indicators divided by the number
of full-time equivalent academic staff (weight of 10\%). The raw data are
normalized by assigning to
the best performing institution a score of 100 and all other institutions
receiving a score relative to the leader. The ARWU score is a weighted
average of the six normalized indicators, which is finally re-scaled to a
maximum of 100. The six indicators have moderate to strong correlations in the
range from 0.48 to 0.87 and an average bivariate correlation of 0.68.

The THES, \citet{THES2008}, summarizes university features related to research
quality, graduate employability, international orientation and teaching
quality using six indicators: the opinion of academics on which institutions
they consider to be the best in the relevant field of expertise (weight of
40\%), the number of papers published and citations received by research staff
(weight of 20\%), the opinion of employers about the universities from which
they would prefer to recruit graduates (weight of 10\%), the percentage of
overseas staff at the university (weight of 5\%), the percentage of overseas
students (weight of 5\%), and finally the ratio between the full-time equivalent
faculty and the number of students enrolled at the university (weight of
20\%). Raw data are standardized. The standardized indicator scores are then
scaled  by dividing by the best score.
The THES score is the weighted average of the six normalized indicators, which is finally
re-scaled to a maximum of 100. The six indicators have very low to
moderate correlations that range from 0.01 to 0.64 and a low
average bivariate correlation of 0.24.

Results for ARWU and THES are given in
Tables~\ref{Table_Si_var-h}-\ref{Table_Si_var-Si} and \ref{table_qmax}.
The first two panels of Table~\ref{Table_Si_var-h} provide the bandwidth selection results
for ARWU and THES; the corresponding panels of Table~\ref{Table_Si_var-Si} give estimates of
the importance measure $S_i$ for different choices of bandwidth. The first two lines in Table~\ref{table_qmax}
give the maximum discrepancy statistic $d_m$ for ARWU and THES. Finally the two upper graphs in Fig.~\ref{fig-qindex}
summarize the comparison between target and actual relative importance of indicators.
For ARWU the main effects $S_{i}$ are more similar to each other than the nominal weights,
i.e. ranging between $0.14$ and $0.19$
(normalised $S_i$ values to unit sum, cross validation estimates)
when weights should either be $0.10$ or $0.20$.

The situation is worse for THES, where the combined importance of
peer review based variables (recruiters and academia) appears larger than
stipulated by developers, indirectly supporting the hypothesis of linguistic
bias at times addressed to this measure (see e.g. \citet{Saisana-al11} for a
review). Further for THES the `teachers to student ratio', a key variable aimed at capturing
the teaching dimension, is much less important than it
should be when comparing normalized $S_{i}$ ($0.09$, cross validation estimate) with the nominal
weight ($0.20$).

Overall, there is more discrepancy between the nominal weights assigned to
the six indicators and their respective main effects in THES ($d_{m,CV}=0.42$) than
in ARWU ($d_{m,CV}=0.36$), cross validation estimates. Comparing this result with the conclusions in \citet{Saisana-al11},
we can see the value-added of the present measure
of importance. In that paper we could not draw a judgement about the
relative quality of THES with respect to ARWU. The main effects used here
allow us to say that -- leaving aside the different normative frameworks
about which no statistical inference can be made -- ARWU is statistically
more consistent with its declared targets than THES.

When considering the sensitivity of $d_m$ values to the choice of bandwidths $h$,
one can see that the range $[d_{m,\min},d_{m,\max}]$ is slightly shorter for ARWU ($[0.26,0.50]$) than for THES
($[0.29,0.55]$); this implies that ARWU is slightly less sensitive than THES to the
choice of bandwidths $h$.
Note however that the two ranges overlap, so that there are choice of
bandwidths $h$ for which the ordering of $d_m$ values is reversed. This, however,
does not happen at the values $h_{CV}$ and $h_{DPI}$.

The hypothesis of linearity is not rejected for two indicators for ARWU
and for four indicators for THES, when evaluating the tests at $h_{DPI}$ and $h_{CV}$.
The two indicators for ARWU are those with the highest proportion
of values equal to 0, which were discarded in the choice of bandwidth;
the number of valid cases $n$ are 198 and 135 respectively.
This may reflect the fact that it is more difficult to reject linearity
with smaller samples. The indicators used in THES instead
do not have so many zero values; also here however, one finds that
$\mathrm{E}(y|x_i)$ is approximately linear for 4 indicators.

\begin{table}
\caption{\label{Table_Si_pil-hi}Bandwidth choice at pillar level.
Bandwidth $h$:
$h_{i,CV}$ (cross validation),
$h_{i,PDI}$ (DPI);
$p$-values for the linearity test:
$p_{i,CV}$ ($p$-value for $h_{i,CV}$),
$p_{i,DPI}$ ($p$-value for $h_{i,DPI}$);
$n$ is the number of observations with $x_{ji}>0$ used for CV and DPI;
$\ast$: right end of the grid $\mathcal{H}$.
}
\centering
\fbox{%
\begin{tabular}{lll lll}
& $h_{CV}$ & $p_{CV}$ & $h_{DPI}$ & $p_{DPI}$ & $n$ \\\hline
\multicolumn{6}{c}{2010 HDI} \\ \hline
Life expectancy               &  0.08    & 0.00    & 0.07    & 0.00    & 169    \\
Education                     &  0.02    & 0.09    & 0.06    & 0.21    & 169    \\
GDP per capita                &  0.05    & 0.00    & 0.06    & 0.00    & 169    \\ \hline
\multicolumn{6}{c}{IAG} \\ \hline
Safety and security           & 17.69    & 0.15    & 3.31    & 0.45    &  53    \\
rule of law and corruption    & ${25.05}^{\ast}$    & 0.30    & 4.75    & 0.94    &  53    \\
part. and human rights        &  4.89    & 0.08    & 2.85    & 0.41    &  53    \\
Sust. economic opportunity    & 22.14    & 0.09    & 4.21    & 0.51    &  53    \\
Human development             & ${25.05}^{\ast}$    & 0.17    & 3.42    & 0.87    &  53    \\
\hline
\multicolumn{6}{c}{SSI} \\ \hline
Personal development          &  0.69    & 0.00    & 0.37    & 0.00    & 151    \\
Healthy environment           & ${25.05}^{\ast}$    & 0.41    & 0.49    & 0.69    & 151    \\
Well-balanced society         &  0.69    & 0.00    & 0.42    & 0.01    & 151    \\
Sustainable use of resources  &  0.30    & 0.00    & 0.30    & 0.00    & 150    \\
Sustainable World             &  0.86    & 0.00    & 0.38    & 0.01    & 151%
\end{tabular}
}
\end{table}

\subsection*{The Human Development Index 2009}
The HDI, see \citet{HDI2009}, summarizes human development in $182$ countries based
on four indicators: a long healthy life measured by life expectancy at birth
(weight of 1/3), knowledge measured by adult literacy rate (weight of 2/9)
and combined primary, secondary and tertiary gross enrollment ratio (weight
of 1/9), and a decent standard of living measured by the GDP per capita
(weight of 1/3). Raw data in the four indicators are normalized by using the
min-max approach to be in {[}0, 1{]}. The 2009 HDI score is the
weighted average of the four normalized indicators. Because data on Adult
literacy rate was missing for several countries, we analyzed data only for the
countries without missing data; this gave a total of 142 countries.
The four indicators present
strong correlations that range from 0.70 to 0.81 and an average bivariate
correlation of 0.74.

Nominal weights and estimates of the main effects are given
in the last panel of Table~\ref{Table_Si_var-Si}, while the choice
of bandwidth is given in Table~\ref{Table_Si_var-h}. The maximum discrepancy
is given in Table~\ref{table_qmax} and a graphical comparison of nominal weights and
estimates of the main effects is provided in Fig.~\ref{fig-qindex}.
Table \ref{Table_Si_var-h} reports evidence on the choice of bandwidth $h$
and the $p$-values for the linearity test, at the values $h_{DPI}$
and $h_{CV}$ of the smoothing parameter $h$.

Both the main effects $S_{i}$ and the Pearson correlation coefficients
reveal a relatively balanced impact of the four indicators
`life expectancy', `GDP per capita', `enrolment in
education', and `adult literacy' on the variance of the HDI scores, with the
adult literacy being slightly less important. It would seem that HDI
depends more equally from its four variables than the weights assigned by the
developers would imply. For example, if one could fix
adult literacy
the variance of the HDI scores would on average be reduced by $77\%$
(CV estimate), whereas by fixing the most influential indicator,
GDP per capita, the
variance reduction would be $84\%$ on average.

One might suspect that it was precisely the developers' intention, when
assigning nominal weights
$11\%$ and $33\%$ to these two variables respectively, to make them
equally important on the basis of the $S_i$ measure; however this is
is not stated explicitly in the index documentation report \citet{HDI2009}.
Overall, there is considerable discrepancy between the nominal weights
assigned to the four indicators and their respective main effects in 2009 HDI ($d_{m,CV}=0.63$).

The analysis of the 2009 HDI illustrates vividly that assigning unequal
weights to the indicators is not a sufficient condition to ensure unequal
importance. Although the 2009 HDI developers assigned weights varying
between $11\%$ and $33\%$, all four indicators are roughly equally
important. The scatterplots in Fig.~\ref{Fig_HDI9-1}-\ref{Fig_HDI9-4} help visualize the
situation. In cases like this, where the variables are strongly and roughly
equally correlated with the overall index, each of them ranks the countries
roughly equally, and the weights are little more than cosmetic.

\subsection{Importance at the pillar level}\label{Case-Studies-pillar}The issue
of weighting is particularly fraught
with normative implications in the case of pillars. As mentioned above,
pillars in composite indicators are often given equal weights on the ground
that each pillar represents an important -- possibly normative -- dimension
which could not and should not be seen to have more or less weight than the
stipulated fraction. The discrepancy measure presented here can be of
particular relevance and interest to gauge the quality of a composite
indicator with respect to this important assumption. Here we consider
the 2010 version of the HDI, the Index of African
Governance (IAG) and the Sustainable Society Index (SSI).
\begin{table}
\caption{\label{Table_Si_pil-Si}Main effects at pillar level.
Nominal weights $w_i$;
main effects $S_i$:
$S_{i,lin}:=S_{i} (\infty)$ (linear fit),
$S_{i,CV}:=S_{i} (h_{CV})$ (cross validation),
$S_{i,DPI}:=S_{i} (h_{DPI})$ (direct plug-in),
$S_{i,\min}:=\min_{h\in \mathcal{H}}S_{i} (h)$,
$S_{i,\max}:=\max_{h\in \mathcal{H}}S_{i} (h)$.}
\centering
\fbox{%
\begin{tabular}{lllllll}
\hline
& $w_{i}$ & $S_{i,lin}$ & $S_{i,CV}$ & $S_{i,DPI}$ & $S_{i,\min }$ & $%
S_{i,\max }$ \\ \hline
\multicolumn{7}{c}{2010 HDI} \\ \hline
Life expectancy               & 0.33    & 0.82    & 0.84    & 0.84    & 0.81    & 0.86    \\
Education                     & 0.33    & 0.86    & 0.87    & 0.86    & 0.84    & 0.89    \\
GDP per capita                & 0.33    & 0.90    & 0.90    & 0.90    & 0.89    & 0.93    \\
\hline
\multicolumn{7}{c}{IAG} \\ \hline
Safety and security           & 0.20    & 0.52    & 0.54    & 0.63    & 0.51    & 0.87    \\
rule of law and corruption    & 0.20    & 0.77    & 0.76    & 0.78    & 0.76    & 1.00    \\
part. and human rights        & 0.20    & 0.44    & 0.63    & 0.68    & 0.43    & 1.00    \\
Sust. economic opportunity    & 0.20    & 0.52    & 0.52    & 0.56    & 0.52    & 0.98    \\
Human development             & 0.20    & 0.50    & 0.50    & 0.55    & 0.49    & 0.94    \\
\hline
\multicolumn{7}{c}{SSI} \\ \hline
Personal development          & 0.13    & 0.05    & 0.14    & 0.17    & 0.04    & 0.27    \\
Healthy environment           & 0.13    & 0.04    & 0.04    & 0.07    & 0.04    & 0.27    \\
Well-balanced society         & 0.13    & 0.13    & 0.21    & 0.21    & 0.12    & 0.32    \\
Sustainable use of resources  & 0.30    & 0.48    & 0.64    & 0.64    & 0.47    & 0.72    \\
Sustainable World             & 0.30    & 0.02    & 0.06    & 0.10    & 0.02    & 0.29  %
\end{tabular}
}
\end{table}

\subsection*{The Human Development Index 2010}\label{sec_sub_HDI2010}
In this section we analyze the 2010 version of the HDI at the pillar level, covering 169 countries. From the methodological
viewpoint the main novelty in this version of the index is the use of a
geometric -- as opposed to an arithmetic -- mean, in the aggregation of the
three pillars. The three pillars cover health (life
expectancy at birth) $x_{\mathrm{life}}$, education $x_{\mathrm{edu}}$ and
income (gross national income per capita) $x_{\mathrm{inc}}$. Education is the combination of two
variables, namely mean years of schooling and expected years of schooling, see
\citet{HDI2010}. The $2010$ HDI\ index $y$ is computed as
\[
y=\left( x_{\mathrm{life}}\cdot x_{\mathrm{edu}}\cdot x_{\mathrm{inc}%
}\right) ^{\frac{1}{3}}
\]
where all three dimensions have equal weights.
The reason for this change of aggregation scheme is to introduce
an element of `imperfect substitutability across all HDI dimensions', i.e.
to reduce the compensatory
nature of the linear
aggregation, see \cite[p. 216]{HDI2010}.

\begin{table}
\caption{\label{table_qmax}Maximum discrepancy statistic $d_m$ for
different choice of the bandwidth $h$ in the main effect estimator
$S_i(h)$. DPI: direct plug-in, CV: cross validation, lin: linear fit,
$\min$ and $\max$ values where obtained by considering all possible
combinations of $S_{i,\ell_i}$ values, where $\ell_i = \min, \max$.}
\centering
\fbox{%
\begin{tabular}{llllll}
 & $d_{m,DPI}$  & $d_{m,CV}$   & $d_{m,lin}$  & $d_{m,\min}$ & $d_{m,\max}$ \\ \hline
ARWU    & 0.36    & 0.36    & 0.31    & 0.26    & 0.50    \\
THES    & 0.41    & 0.42    & 0.34    & 0.29    & 0.55    \\
2009 HDI & 0.59    & 0.63    & 0.57    & 0.50    & 0.69    \\
\hline
2010 HDI & 0.06    & 0.07    & 0.09    & 0.03    & 0.13    \\
IAG     & 0.29    & 0.34    & 0.42    & 0.13    & 0.57    \\
SSI     & 0.85    & 0.91    & 0.95    & 0.38    & 0.98   %
\end{tabular}}
\end{table}

Nominal weights and estimates of the main effects are given in
the first panel of Table~\ref{Table_Si_pil-Si}, while the choice
of bandwidth is given in Table~\ref{Table_Si_pil-hi}. The maximum discrepancy
is given in row 4 of Table~\ref{table_qmax} and a graphical comparison of nominal weights and
estimates of the main effects is provided in Fig.~\ref{fig-qindex}.

Overall, the HDI 2010 shows very little discrepancy between
the goals of equal importance of the three pillars and the main effects. In fact
all three pillars have similar impact on the index variance
(roughly $84-90\%$). Hence, in this case
the relative nominal weights are approximately equal to the relative
impact of the pillars' on the index variance. Such a correspondence
is of value because it indicates than no pillar impacts too much or too
little the variance of the index as compared to its `declared' equal importance.
Compared to the other examples discussed, the 2010 HDI is the
most consistent in this respect ($d_{m,CV}=0.07$).
The linearity tests reveal that the role of education is approximately linear within the index,
despite the multiplicative aggregation scheme.

In order to assess the impact of the choice of the aggregation scheme on the index balance,
we also perform a counterfactual analysis of the 2010 HDI using linear aggregation of the
three dimensions.
We find that this choice does not affect the relative importance of dimensions,
as these have comparable variances and covariances.
Hence the 2010 HDI would have been balanced also under a linear aggregation scheme.
This, however, does not detract from the conceptual appeal of imperfect substitutability
implicit in geometric aggregation.

\subsection*{Index of African Governance}
The Index of African Governance was developed by the Harvard Kennedy School, see
\citet{RotbergGiss2008}; for a validation study see \citet{Saisa-IAG}.
In the 2008 version of the index, 48 African countries are ranked according to five-pillars: (i)
Safety and Security, (ii) Rule of Law, Transparency, and Corruption, (iii)
Participation and Human Rights, (iv) Sustainable Economic Opportunity, and
(v) Human Development. The five pillars are described by fourteen
sub-pillars that are in turn composed of $57$ indicators in total (in a
mixture of qualitative and quantitative variables). Raw indicator data were
normalized using the min-max method on a scale from $0$ to $100$. The five
pillar scores per country were calculated as the simple average of the
normalized indicators. Finally, the IAG scores were
calculated as the simple average of the five pillar scores. The five pillars
have correlations that range from 0.096 to 0.76 and average bivariate
correlation of 0.45. Three pairwise correlations (involving Participation
and Human Rights and either Sustainable Economic Opportunity or Human
Development or Safety \& Security) are not statistically significant at the
5\% level.

\begin{figure}
\centering
\makebox{\includegraphics[scale=0.75]{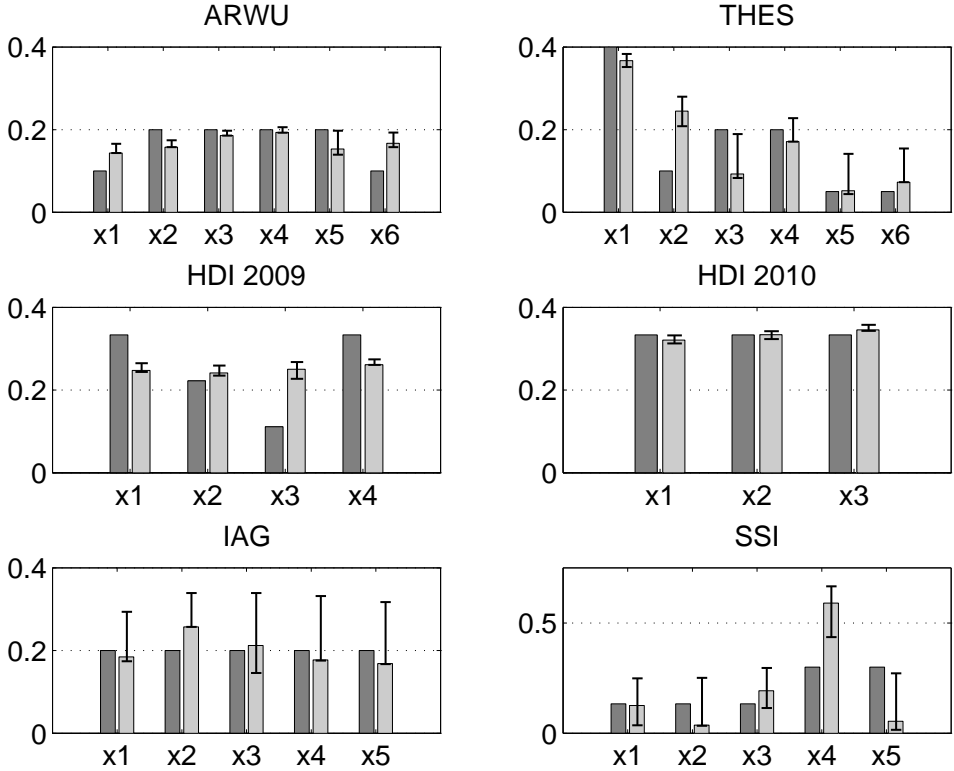}}
\caption{\label{fig-qindex}Comparison of normalized main
effects $S_{i}^\ast$  (in light grey) and $w_i$ (in dark grey).
$S_{i}^\ast :=S_{i} / c$ with $c:=\sum_{t=1}^k S_{t}$ and
$S_i:=S_{i,CV}$ (cross validation).
The indicators $x_i$ are numbered consecutively as in
Tables~\ref{Table_Si_var-h}--\ref{Table_Si_pil-Si}.
Bounds for $S_{i}^\ast$ are constructed
as $S_{i,\min} / c$, $S_{i,\max} / c$.
}
\end{figure}

Nominal weights and main effects are given
in Table \ref{Table_Si_pil-Si} and in Fig.~\ref{fig-qindex}, while the
choice of bandwidth is reported in Table \ref{Table_Si_pil-hi} and
the discrepancy statistics in Table \ref{table_qmax}. The main conclusions are
summarized as follows: The IAG is a good example of
the situation discussed in Section \ref{sec_Introd} whereby all pillars
represent important normative elements which by design should be equally
important in the developers' intention. Overall the IAG appears to be balanced with respect to four pillars that have similar
impact on the index variance (roughly $50-63\%$), but the fifth pillar on the Rule of Law is more
influential than conceptualised ($S_i=76\%$, cross validation estimate).
The IAG has a maximal discrepancy statistic $d_{m,CV}=0.34$.

The linearity tests in Table \ref{Table_Si_pil-hi} suggest that there is
no statistical evidence against linearity for all the five indicators. Hence one could
calculate $S_i$ here as $R_i^2$.

\subsection*{Sustainable Society Index}
The Sustainable Society Index (SSI) has been developed by the Sustainable
Society Foundation for 151 countries and it is based on a
definition of sustainability of the Brundtland Commission \citep%
{vdKerkManuel2008}. Also in this example, the five pillars of the index
represent normative dimensions which are, however, considered of different
importance: Personal Development (weight of 1/7), Healthy Environment (1/7),
Well-balanced Society (1/7), Sustainable Use of Resources (2/7), and
Sustainable World (2/7). These five pillars are described by $22$
indicators. Raw indicator data were normalized using the min-max method on a
scale from 0 to 10. The five pillars were calculated as the simple average
of the normalized indicators. The SSI scores were calculated as the weighted
average of the five pillar scores.

One can note that the linearity test suggests that for the second pillar
`Healthy Environment' there is no evidence against linearity of its relation to the SSI index.
The five pillars have correlations that
range from $-0.62$ to $0.75$,
where negative correlations between pillars are generally undesired,
as they suggest the presence of trade-offs between pillars
(e.g. economic performance can only come with an environmental cost).
Such trade-offs within index dimensions are a reminder of the
danger of compensability between dimensions.

For the Sustainable Society Index, there are notable differences between
declared and variance-based importance for the five pillars. The
different association between a pillar and the overall index can also be
grasped visually in Fig.~\ref{fig-qindex}.
The two pillars on `Sustainable use of resources' and on `Sustainable World'
are meant to be equally important accordingly to the nominal weights (2/7
each), while the main effects suggest that the variance reduction obtained
by fixing the former is $67\%$ compared to merely $9\%$ by fixing the
latter. This strong discrepancy is due to the significant negative
correlations present among the SSI pillars. Overall, the level of maximal
discrepancy of the SSI is the highest of the examples discussed ($%
d_{m,CV}=0.91$).
The authors and the developers of the SSI have been communicating on this issue,
and the 2010 version of the SSI index appears considerably improved,
see \texttt{http://www.ssfindex.com/ssi/}.

\subsection{Reverse-engineering the weights}\label{Correcting-results}Applying
the reverse engineering
exercise described in Section \ref{Correcting} and the Appendix
to our test cases
(except for the case of 2010 HDI that has low maximal discrepancy
between relative weights and relative importance for the three pillars, and it is
not obtained by the linear aggregation scheme (\ref{def})),
we find that to achieve a relative impact of the indicators (or pillars) (as
measured by the square of the Pearson correlation coefficient $R_{i}^{2}$)
that equals the relative  `declared'\ importance of the indicators,
negative nominal weights are involved in all studies except for the SSI. In
the case of SSI, to guarantee that the two pillars on
Sustainable Use of Resources and Sustainable World are twice as important as
the other three pillars, the nominal weights to be assigned to them are
Personal Development (weight of 0.19), Healthy Environment (0.16),
Well-balanced Society (0.07), Sustainable Use of Resources (0.16), and
Sustainable World (0.41).  For all other cases,
the data correlation structure does not allow the developers to achieve
the stated relative importance by choosing positive weights.

\section{Conclusions}
\label{Conclusions}
According to many -- including some of the authors of the
Stiglitz report, see \citet{Stiglitz2009} -- composite indicators have serious
shortcomings. The debate among those who prize their pragmatic nature
in relation to pragmatic problems, see
\citet{Hand2009}, and those who consider them an aberration is unlikely to be
settled soon, see \citet{Saltelli-2007} for a review of pros and cons.
Still these measures are pervasive in the
public discourse and represent perhaps the best known face of statistics
in the eyes of the general public and media.

One might muse that what official statistics are to the consolidation of the
modern nation state, see \citet{Hacking1990}, composite indicators are to the
emergence of post-modernity, -- meaning by this the philosophical critique
of the exact Science and rational knowledge programme of Descartes and
Galileo, see \citet{Toulmin90}, p. 11-12. On a practical level, it is undeniable
that composite indicators give voice to a plurality of different actors and
normative views. 
The authors in \citet{Stiglitz2009} remark (p. 65):\\
\\
\emph{ ``The second {[}argument against composite indicators{]} is
a general criticism that is frequently addressed at composite indicators,
i.e. the arbitrary character of the procedures used to weight their various
components. (...) The problem is not that these weighting procedures are
hidden, non-transparent or non-replicable --- they are often very explicitly
presented by the authors of the indices, and this is one of the strengths of
this literature. The problem is rather that their normative implications are
seldom made explicit or justified.''}
\\
\\
The analysis of this paper shows that,
although the weighting
procedures are often very explicitly presented by the authors of the
indices, the implications of these are neither fully understood, nor
assessed in relation to the normative implications.
This paper proposes a variance-based tool to measure the internal discrepancy
of a composite indicator between target and effective importance.

Our main conclusions can be summarized as follows.
For transparency and simplicity, composite indicators are most often built using
linear aggregation procedures which are fraught with the
difficulties described in the Introduction: practitioners know that weights
cannot be used as importance, while they are precisely elicited as if they
were. Weights are instead measures of substitutability in linear aggregation. The
error is particularly severe when a variable's weight substantially deviates
from its relative strength in determining the ordering of the units (e.g.
countries) being measured.

Pearson's correlation ratio (or main effect) that is suggested in this paper is a suitable
measure of importance of a variable (be it indicator or pillar) because:
\textit{i}) it offers a precise definition of importance (that is `the expected reduction in variance of the
composite indicator that would be obtained if a variable could be fixed'), \textit{ii})
it can be used regardless of the degree of correlation between variables, \textit{iii})
it is model-free, in that it can be applied also in non-linear aggregations,
and finally \textit{iv}) it is not invasive, in that no changes are made to the
composite indicator or to the correlation structure of the indicators.

Because of property \textit{i}) and the fact that
it takes the whole covariance structure into account,
the main effect can also be useful to prioritise variables on which a country
or university, or whatever units are being
rated,
could intervene to improve its overall score.
Note that the indicator with highest main effect
is not necessarily the one in which the country scores the worst.

The main effects approach can complement the techniques for robustness
analysis applied to composite indicators thus far seen in the literature,
see e.g. \citet{Saisana-al05, OECD2008, Saisana-al11}.
The approach described in this paper does not need an explicit modeling
of error-propagation but it is simply based on the data as produced by developers.

The discrepancy statistic based on the absolute error
between ratios of the main effects and of the corresponding target relative
importance provides a pragmatic answer to the research question posed in this paper.
Relative main effects are variance-based,
and hence they are ratios of quadratic forms of nominal weights,
while target relative importance are often deduced as ratios of nominal weights.
Comparing them via the discrepancy statistic
is a way to compare these two importance measures, one of which is stated ex-ante as a target
and the other one that is computed ex-post; this allows to see how close the two measures are in practice.

The discrepancy statistic has been effective in the six examples
discussed, in that it allowed an analytic judgement about the discrepancy in
the assignment of the weights in two well known measures of higher education
performance ($d_{m}=0.42$ for THES versus
$d_{m}=0.36$ for
ARWU),
two versions of a human development index
($d_{m}=0.63$ for the 2009 HDI and $d_{m}=0.02$ for the 2010 HDI),
one index of governance
($d_{m}=0.34$ for the IAG)
and one index of sustainability
($d_{m}=0.86$ for the SSI).

Our reverse engineering analysis shows that in most cases it is not possible
to find nominal weights that would give the desired importance to variables.
This can be a useful piece of information to developers, and might induce a
deeper reflection on the cost of the simplification achieved with linear
aggregation. Developers could thus:
\begin{enumerate}
\item[a)]
avoid associating nominal weights with importance, but inform users of the
relative importance of the variables or pillars, using statistics such as those
presented in this paper;
\item[b)] abstain from aggregating pillars when these display important trade offs
which make it difficult to give them target weights in an aggregated
index;
\item[c)] reconsider the aggregation scheme, moving from the linear one (which is fully
compensatory) to a partially- or fully-non-compensatory alternative, such as e.g.
a Condorcet-like (or approximate Condorcet) approach, where weights would
fully play their role as measure of importance, see \citet{Munda2008};
\item[d)] assess different weighting strategies, so as to select the one that leads to a
minimum discrepancy statistic between target weights and variables importance.
\end{enumerate}

\section*{Acknowledgement}
We thank, without implicating, Beatrice d'Hombres, Giuseppe Munda,
the Associate Editor and two Referees for useful comments. The views expressed
are those of the authors and not of the European Commission or the University of Insubria.

\section*{Appendix - Solution to the inverse problem}
\label{INV}In the linear case, the ratio $S_i/ S_1$ equals the ratio of squares of
Pearson's correlation coefficients $R_{i}^{2}/R_{1}^{2}$;
this is a function $H_{i}\left( \mathbf{w}\right) $ of $\mathbf{w}:=(w_{1},\dots ,w_{k})$
and of the covariance matrix $\mathbf{\Sigma }$ of $\mathbf{x}:=(x_{1},\dots
,x_{k})^{\prime }$. One finds $H\left( \mathbf{w}\right) =\left( \mathbf{e}%
_{i}^{\prime }\mathbf{\Sigma w}\right) ^{2}\sigma _{11}/(\left( \mathbf{e}%
_{1}^{\prime }\mathbf{\Sigma w}\right) ^{2}\sigma _{ii})$, where $\mathbf{e}%
_{i}$ is the $i$-th column of the identity matrix of order $k$ and $\sigma
_{ii}$ is the $i$-th variance on the diagonal of $\mathbf{\Sigma }$. We
wish to make $H_{i}\left( \mathbf{w}\right) $ equal to a pre-selected
value $z _{i}^{2}$ for all $i$:\
\begin{equation}
H_{i}\left( \mathbf{w}\right) =z _{i}^{2},\qquad i=1,\dots ,k,
\label{eq_zeta=z}
\end{equation}%
and seeks to find a solution $\mathbf{w}\in \mathbb{R}^{k}$ to this problem
such that nominal weight to sum to 1, i.e.
\begin{equation}
\mathbf{1}^{\prime }\mathbf{w}=1.  \label{eq_constraint}
\end{equation}%
We show that this solution is unique and it is given by
(\ref{eq_w_inverse_problem}) in the text, where
 $\mathbf{g}$ is a vector with $i$-th entry equal to
 $g_{i}:=z_{i}\sqrt{\sigma _{ii}/\sigma _{11}}>0$ and $\mathbf{1}$
 is a $k$-vectors of ones.

Note that by construction $g_{1}=1$. One has that (\ref{eq_zeta=z}) can be
written as $\mathbf{e}_{1}^{\prime }\mathbf{\Sigma w}-\frac{1}{g_{i}}\mathbf{%
e}_{i}^{\prime }\mathbf{\Sigma w}=\mathbf{0}$, or, setting $\mathbf{G}:=%
\mathrm{diag}(1,1/g_{2},\dots 1/g_{k})$, $\mathbf{F}:=\mathbf{1e}%
_{1}^{\prime }$ as $\left( \mathbf{F-G}\right) \mathbf{\Sigma w=0}$. This
shows that $\mathbf{\Sigma w}$ should be selected in the right null space of
$\mathbf{F-G}$.
We observe that
\[
\mathbf{F-G=}\left(
\begin{array}{cccc}
0 & 0 &  & 0 \\
1 & -1/g_{2} &  & \vdots \\
\vdots &  & \ddots & 0 \\
1 & 0 &  & -1/g_{k}%
\end{array}%
\right)
\]
whose right null-space $\mathcal{A}$ is one dimensional;
moreover $\mathcal{A}$ is spanned by $\mathbf{g}%
:=(1,g_{2},\dots ,g_{k})^{\prime }$. Hence $\mathbf{\Sigma w=g}c$ for a
nonzero $c$ or $\mathbf{w=\Sigma }^{-1}\mathbf{g}c$. Substituting this
expression in (\ref{eq_constraint}), one finds $1=\mathbf{1}^{\prime }%
\mathbf{w=1}^{\prime }\mathbf{\Sigma }^{-1}\mathbf{g}c$, which implies $c=1/%
\mathbf{1}^{\prime }\mathbf{\Sigma }^{-1}\mathbf{g}$. One hence concludes
that the weights that satisfy (\ref{eq_zeta=z}) are given by
(\ref{eq_w_inverse_problem}), and that they are unique.


\end{document}